# Towards Quantum-Safe O-RAN: Experimental Evaluation of ML-KEM-Based IPsec on the E2 Interface

Mario Perera, Michael Mackay, Max Hashem Eiza, Alessandro Raschellà, Nathan Shone, and Mukesh Kumar Maheshwari

***Abstract*—As Open Radio Access Network (O-RAN) deployments expand and adversaries adopt "store-now, decrypt-later" strategies, operators need empirical data on the cost of migrating critical control interfaces to post-quantum cryptography (PQC). This paper experimentally evaluates the impact of integrating a NIST-aligned Module-Mattice Key-Encapsulation Mechanism (ML-KEM) into IKEv2/IPsec, protecting the E2 interface between the 5G Node B (gNB) and the Near-Real-Time RAN Intelligent Controller (Near-RT RIC). Using an open-source testbed built from srsRAN, Open5GS, FlexRIC and strongSwan (with liboqs), we compare three configurations: no IPsec, classical Elliptic Curve Diffie–Hellman (ECDH)-based IPsec, and ML-KEM-based IPsec. The study focuses on IPsec tunnel-setup latency and the runtime behaviour of Near-RT RIC xApps under realistic signalling workloads. Results from repeated, automated runs show that ML-KEM integration adds a small overhead to tunnel establishment, which is approximately 3~5 ms in comparison to classical IPsec, while xApp operation and RIC control loops remain stable in our experiments. These findings, produced from an open, reproducible testbed, indicate that ML-KEM–based IPsec on the E2 interface is practically feasible and inform quantum-safe migration strategies for O-RAN deployments.**

***Index Terms*—5G Network, Post-Quantum Cryptography, OpenRAN, O-RAN, IPSec, ML-KEM**

## I. Introduction

Secure communication protocols are fundamental to trustworthy mobile networks in increasingly complex Fifth Generation (5G) deployments. Open Radio Access Networks (O-RANs), aka Open-RANs, have emerged as a key architectural paradigm for 5G and beyond, enabling service heterogeneity, multi-vendor interoperability, virtualisation, and AI-driven intelligence. Specifically, O-RAN defines a fully virtualised RAN in which base station functions are disaggregated into an open Central Unit (CU) and Distributed Unit (DU) and deployed within a cloud environment or data centre, along with a hardware-based Radio Unit (RU) interconnected through open interfaces. These components are managed and optimised by centralised control entities, namely the RAN Intelligence Controllers (RICs), which maintain a comprehensive view of the network [1]. This disaggregation, however, significantly enlarges the attack surface, making interface protection a central security concern [2].

Therefore, the O-RAN Alliance Working Group (WG11), has positioned "security by design" as a core principle, specifying mandatory controls such as TLS and IPsec for inter-node communication [3]. Among these, the E2 interface linking the 5G Node B (gNB)'s CU/DU with the Near-RT RIC is particularly critical as it carries near-real-time control and telemetry data and is subject to stringent timing constraints. WG11 mandates IPsec on E2 to ensure confidentiality and integrity, but stronger cryptographic mechanisms may impact latency and resource usage in ways that are not yet fully explored [3, 4].

In parallel, advances in quantum computing threaten widely deployed public-key primitives such as Rivest–Shamir–Adleman (RSA), Elliptic Curve Diffie–Hellman (EC)DH, and Elliptic Curve Cryptography (ECC) thus, the long-term security of IPsec and TLS. This has motivated intensive work on Post-Quantum Cryptography (PQC) and the standardisation of quantum-resistant schemes. The National Institute of Standards and Technology (NIST) has recently finalised FIPS 203, specifying the Module-Lattice-based Key-Encapsulation Mechanism (ML-KEM, formerly CRYSTALS-Kyber) as a general-purpose KEM for quantum-safe key establishment, with parameter sets ML-KEM-512/768/1024 that trade performance against security [5]. Integrating such primitives into O-RAN interface security is essential to mitigate "store-now, decrypt-later" attacks over the lifetime of 5G and future sixth generation (6G) systems.

Existing work has begun to address both sides of this problem. On the O-RAN side, several studies analyse architectural security risks, interface threats, and mitigation strategies, and provide comprehensive reviews of O-RAN security challenges and opportunities (e.g., [6]). Other works quantify the impact of classical encryption on O-RAN interfaces, including E2 and the Open Fronthaul, using testbeds and network emulators (e.g., [7]). On the PQC side, there are empirical evaluations of PQC KEMs in 5G core networks (e.g., PQC-enabled TLS in free5GC) and in other wireless contexts, focusing mainly on handshake latency and message sizes rather than RAN control interfaces [8]. More recently, work on introducing PQC into O-RAN has considered integrating post-quantum KEMs into IPsec and MACsec to protect selected O-RAN interfaces but focuses primarily on design and feasibility across multiple links rather than detailed, per-interface performance analysis [9].

Despite this progress, a critical gap remains in which there is limited empirical evidence on how ML-KEM–based key encapsulation, when integrated into IPsec in line with O-RAN WG11 guidance, affects the performance of the E2 interface in realistic O-RAN deployments. Particularly, network operators lack open, reproducible testbeds and quantitative measurements that characterise the trade-off between quantum-resistant security and near-real-time responsiveness for RIC-driven control.

Against this background, this paper investigates whether ML-KEM–based post-quantum key encapsulation imposes significant performance issues when used to protect the E2 interface of 5G O-RAN via IPsec. The overall aim is to model and experimentally evaluate the performance impact of PQC, specifically ML-KEM, on an IPsec-protected E2 interface in a 5G O-RAN deployment. To achieve this aim, the paper pursues three objectives. Firstly, review 5G O-RAN architectures, security requirements of O-RAN interfaces, and PQC algorithms, and focus on ML-KEM as a suitable KEM candidate for securing E2 in accordance with emerging standards. Secondly, design and deploy an open-source, reproducible experimental platform emulating a 5G O-RAN with Near-RT RIC, using the following components: srsRAN [10], Open5GS [11], FlexRIC [12] and strongSwan with ML-KEM support [13] to realise IPsec-based E2 protection. Finally, quantify the impact of ML-KEM–based IKEv2 on the E2 interface, comparing 1) no IPsec; 2) classical ECDH-based IPsec, and 3) ML-KEM–based IPsec in terms of tunnel setup latency, and the runtime behaviour of Near-RT RIC xApps.

Hence, the novel contributions of this paper are threefold: 1) it presents, to the best of our knowledge, the first empirical evaluation of ML-KEM–based IPsec on the O-RAN E2 interface using an open, standards-aligned 5G O-RAN testbed; 2) it provides a documented, reproducible framework for studying PQC-enhanced interface security in O-RAN; and 3) it delivers quantitative evidence that ML-KEM integration primarily increases IPsec tunnel setup latency by a few milliseconds while introducing no observable disruption to Near-RT RIC or xApp operation, thereby informing quantum-safe migration strategies for O-RAN deployments.

The rest of this paper is organised as follows. Section II presents a literature review covering the O-RAN architecture and interface security, prior work on PQC integration in network protocols, and recent industry and standards activity relevant to quantum-safe RANs. Section III describes the testbed design and measurement methodology, including component selection, topology, IPsec/IKEv2 profiles, and the signalling workloads used to exercise the E2 interface. Section IV details the implementation of the testbed and its configuration. Section V reports the evaluation results and analysis of tunnel-setup latency, xApp runtime behaviour under the three security modes, and discussion of operational implications. Finally, Section VI concludes and summarises the main findings and introduces ideas for future work.

## II. Literature Review

This literature review introduces O-RAN architecture including its interfaces and security, and the latest works on PQC evaluation in network protocols (IPsec/TLS), and recent industry and standards activity towards quantum-safe telecoms.

### *A. O-RAN Architecture*

The O-RAN Alliance rearchitects the 5G RAN by extending the 3GPP NG-RAN “split” gNB into a fully disaggregated, cloud-native, and programmable platform. Building on the logical separation of the gNB into CU, DU, and RU in 3GPP Release 15, O-RAN renames these elements O-CU, O-DU, and O-RU, respectively, and standardises open, interoperable interfaces between them. This approach inherits ideas from Cloud (C-RAN) and Virtualised RAN (V-RAN) and goes further by mandating openness of interfaces and embedding intelligence as a first-class design goal [14]. It is now the reference architecture for most academic and industrial work on Open RAN [15]. The O-RAN architecture proposed by the O-RAN Alliance is depicted in Fig. 1.

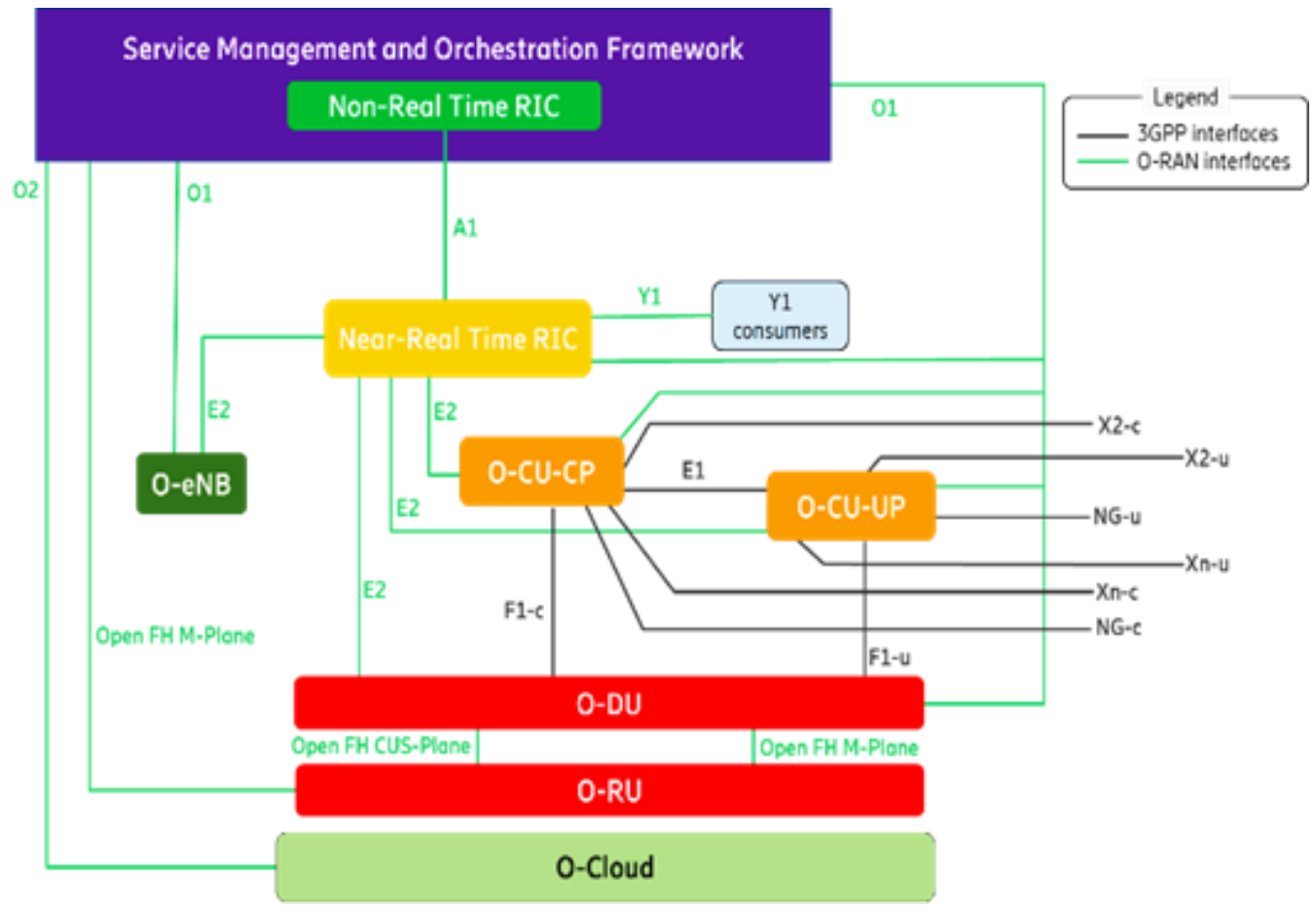

Figure 1: Logical Architecture of O-RAN [16]

At the radio side, the O-CU serves as the intelligent control hub that acts as the bridge between the RAN and the core network. It handles higher-layer protocol processing up to Packet Data Convergence Protocol (PDCP), terminates the NG interface towards the 5G core, and coordinates with the O-DU over the 3GPP F1-C/F1-U interfaces. The O-DU performs mid-layer RAN functions such as Radio Link Control (RLC), MAC, and upper PHY, and aggregates multiple O-RUs via the Open Fronthaul interface. The O-RU is responsible for Radio Frequency (RF) processing and lower PHY, positioned close to the antenna to meet tight timing constraints while leveraging the O-DU for more compute-intensive baseband tasks. These virtualised network functions (O-CU, O-DU, and in many deployments the Near-RT RIC) execute on an O-Cloud infrastructure, which provides the underlying compute, storage, and networking platform for O-RAN Virtual Network Functions (VNFs) and Cloud Native Functions (CNFs).

On the management and control side, the Service Management and Orchestration (SMO) framework acts as the central management plane for the RAN domain. It consolidates traditional Operations Support Systems (OSS) and Network Management Systems (NMS) functions with cloud-native lifecycle management and coordinates RAN optimisation via the RIC. The RIC is split into a Non-Real-Time RIC (Non-RT RIC) operating at time scales above 1s and hosted within the SMO, and a Near-Real-Time RIC (Near-RT RIC) targeting 10ms–1s control loops for Radio Resource Management (RRM) and policy enforcement. Non-RT RIC rApps perform long-term optimisation, analytics, and model training, while Near-RT RIC xApps execute fine-grained control actions such as load balancing, interference management, and radio-bearer optimisation in close interaction with O-CU/O-DU. This decoupled, application-centric control framework is one of the key architectural differentiators of O-RAN.

These components are interconnected through a set of standardised open interfaces that form the backbone of O-RAN's interoperability. The A1 interface links the Non-RT RIC (in the SMO) to the Near-RT RIC, carrying high-level policies, ML models, and aggregated KPIs for policy-driven optimisation. The E2 interface connects the Near-RT RIC to O-CU and O-DU, exposing near-real-time telemetry and control data that enables xApps to implement closed-loop optimisation while respecting RAN timing constraints. The O1 interface provides Fault, Configuration, Accounting, Performance, and Security (FCAPS) management between the SMO and O-RAN nodes (O-CU, O-DU, O-RU, and Near-RT RIC), whereas O2 connects the SMO to the O-Cloud for orchestration of virtualised RAN resources. At the fronthaul, the Open Fronthaul (O-FH) interface between O-DU and O-RU is specified with distinct control, user, synchronisation, and management planes to support multi-vendor interoperability and flexible deployment options. Collectively, these components and interfaces realise a modular, cloud-native RAN that is open to multi-vendor implementations and programmable control, while also introducing new interface surfaces that must be secured in practice.

### *B. O-RAN Security & Post–Quantum Cryptography (PQC)*

The openness and disaggregation of O-RAN, virtualisation of RAN functions on O-Cloud infrastructure, the introduction of programmable RIC platforms running xApps/rApps, and the exposure of multiple interfaces, all create additional points where adversaries can target control and management traffic. The increased exposure covers both passive and active threats. For instance, adversaries can perform passive eavesdropping (e.g., sniffing), active man-in-the-middle or packet-injection attacks that tamper with control messages, protocol-level manipulations that subvert RIC decisions, and resource-exhaustion attacks (e.g., Denial-of-Service (DoS)). Threat modelling studies and recent security reports such as [17] consistently highlight risks such as DoS against control-plane interfaces, zero-day exploits, supply-chain compromise of software components, and attacks on AI/ML pipelines within the RIC. In particular, the E2 interface, which carries near-real-time control and telemetry data between the Near-RT RIC and O-CU/O-DU, is recognised as highly sensitive. Successful attacks on the E2 can directly degrade radio-resource management, QoS, and stability of deployed xApps.

To address these risks, the O-RAN Alliance Security WG11 has defined a set of mandatory security controls per interface, specifying the use of standardised cryptographic protocols to provide confidentiality, integrity, entity authentication, and replay protection [18]. In the current framework, TLS (1.2/1.3) is mandated for many management and policy interfaces (e.g., O1, O2, A1 in Figure 1), MACsec or IPsec are recommended for the Open Fronthaul, and IPsec is mandated for E2, where tunnel-based protection aligns with the need for strong, end-to-end security between logically separate domains. From a protocol-design viewpoint, these controls are realised by combining symmetric ciphers, asymmetric key exchange and digital signatures, and cryptographic hash functions within TLS and IPsec handshakes. While TLS is widely used for secure client–server and management communications at the transport layer, IPsec operates at the network layer to protect inter-component tunnels such as E2 and is the focus of this work.

Furthermore, the O-RAN next Generation Research Group (nGRG) stresses that protecting O-RAN deployments requires moving beyond interface encryption to platform-level assurances [17]. It advocates trusted-computing techniques backed by hardware roots of trust (e.g., Trusted Platform Module (TPM), CPU enclaves) and layered remote attestation (e.g., IETF Remote ATtestation procedureS (RATS) and Trusted Computing Group Device Identifier Composition Engine (DICE)) to establish and maintain trust in O-RAN components, O-Cloud hosts, and management systems. In this view, a trusted O-RAN is one where O-RU/O-DU/O-CU, RIC platforms, and cloud infrastructure can cryptographically prove identity and integrity at boot and runtime, enabling zero-trust enforcement decisions across the architecture [19]. This platform-security perspective complements the protocol-centric view adopted by WG11 and reinforces the need to ensure that PQC-enabled IPsec and TLS can run on these trustworthy platforms.

However, the long-term security of these TLS and IPsec deployments is threatened by advances in quantum computing. Both protocols typically rely on public-key primitives such as RSA, Diffie–Hellman, and ECC for key establishment and authentication. These primitives can be solved efficiently by Shor's algorithm, while Grover's algorithm weakens the effective security of symmetric ciphers. As noted by the O-RAN nGRG report [20], although current quantum hardware cannot yet break deployed schemes, the "harvest-now, decrypt-later" threat is already relevant because mobile network traffic often has long confidentiality lifetimes and infrastructure generations last 15–20 years or more. This motivates an early transition to quantum-safe mechanisms that can be integrated into existing protocol frameworks such as TLS and IPsec without requiring quantum hardware.

PQC provides such mechanisms that can resist known quantum attacks while remaining implementable on classical hardware. NIST's multi-year standardisation process has now culminated in the publication of FIPS 203, 204, and 205, which specify ML-KEM and related signature schemes as the first generation of PQC standards [21]. ML-KEM offers three parameter sets (ML-KEM-512/768/1024) that trade off security strength and performance and is designed specifically for use as a drop-in replacement for classical KEMs in protocols such as TLS and IKEv2. Surveys and position papers (e.g., [22]) on PQC for networks emphasise that, given the long lead times historically observed when transitioning from one public-key paradigm to another (e.g., from RSA to ECC), experimentation and staged deployment must begin well before quantum adversaries become practical.

Recent research has begun to explore how PQC can be integrated into the security protocols relevant to 5G and beyond. Scalise *et al*. [8] extended the free5GC core network with PQC KEMs for TLS 1.3, showing modest increases in handshake time and packet size but no prohibitive overheads in VNF-to-VNF signalling. García *et al*. [23] proposed hybrid and triple-hybrid TLS 1.3 schemes that combine PQC, classical cryptography, and Quantum Key Distribution (QKD) to achieve quantum-resistant authenticated key exchange while maintaining practical performance. Rathi *et al*. [24] presented Q-RAN, a full-stack PQC framework for O-RAN. It integrates ML-KEM for encryption and ML-DSA for signatures and quantum entropy into all control-plane protocols. Specifically, PQ-IPsec, PQ-DTLS and PQ-mTLS are deployed on every O-RAN interface, anchored by a centralised post-quantum certificate authority. They showed it is feasible to retrofit O-RAN with hybrid PQC handshakes (e.g. TLS/IPsec) using ML-KEM and ML-DSA while retaining compatibility. Other studies provide generic performance evaluation frameworks for PQC-enabled TLS, enabling systematic comparison of different NIST-selected KEMs and signature schemes [25].

For IPsec and lower-layer protocols, Bae *et al*. [26] evaluated PQC-integrated IKEv2 in IPsec, quantifying the overheads of different KEMs in tunnel establishment and showing that, although PQC increases computational load and handshake time, the impact can be acceptable in many deployment scenarios. Most recently, a study on introducing PQC algorithms in O-RAN interfaces demonstrated how PQC KEMs can be integrated into IPsec to protect Open Fronthaul links, using strongSwan and liboqs and reporting throughput, delay, jitter, and resource usage for multiple KEM choices [27]. That work confirms both the feasibility of PQC-enabled IPsec in O-RAN context and the importance of carefully characterising performance trade-offs. The O-RAN nGRG complements these efforts by mapping PQC use cases across 3GPP and O-RAN interfaces including E2 and outlining migration strategies such as crypto-agile designs and staged deployment of PQC and QKD [20].

### *C. PQC for O-RAN – Latest Industrial & Standardisation Efforts*

A recent collaboration between PQC vendor Patero and O-RAN vendor Eridan demonstrated Patero's CryptoQoR PQC suite running on Eridan's 5G radios [28]. They set up a private 5G Open RAN network in which Patero's post-quantum encryption protected the data path. The demo showed end-to-end quantum-resistant 5G communication using cryptographic algorithms aligned with the new NIST PQC standards. This proof-of-concept highlights that commercial O-RAN hardware can support PQC without changing the radio stack, providing a quantum-safe solution for critical infrastructure.

In 2024, Nokia and Turkcell successfully trialled a quantum-safe mobile transport link [29]. In their demo, Nokia's IPsec Security Gateway in Turkcell's network was configured with post-quantum algorithms (NIST ML-KEM, ML-DSA, etc.) to secure subscriber data. The partners reported a "world-first" implementation of quantum-safe IPsec in a live 5G mobile network, effectively protecting current traffic against future quantum attacks. While this test was on the transport network rather than a disaggregated O-RAN interface, it demonstrates the viability of PQC-IPsec in telecom networks and sets a precedent for extending such solutions into O-RAN deployments.

In the 3GPP ecosystem, studies have begun on migrating cellular cryptography to PQC [30]. 3GPP TR 33.938 provides a cryptography inventory for 5G/6G and outlines how NIST PQC algorithms might replace existing keys. Future 3GPP work (SA3 security specifications) will evaluate hybrid and pure-PQC handshakes for procedures like Subscription Concealed Identifier (SUCI) concealment and TLS/IKE tunnels.

Despite this growing body of work, an analytic gap remains at the intersection of PQC and O-RAN interface security. While recent work has explored PQC-enabled IPsec in O-RAN, its focus has been primarily on the Open Fronthaul and hybrid key exchange rather than on fully ML-KEM-based key establishment for Near-RT control. In this context, this work positions itself as a focused contribution on quantum-safe O-RAN security. Building on the cryptographic and protocol background above and leveraging NIST-standardised ML-KEM and PQC-enabled IPsec implementation strongSwan, it provides an empirical evaluation of ML-KEM–based key encapsulation for IPsec on the 5G E2 interface. By quantifying the impact on tunnel setup latency, and the behaviour of Near-RT RIC/xApps, the work addresses a key open question for operators: whether quantum-safe protection of critical O-RAN control interfaces can be achieved without compromising stringent performance requirements.

## III. Testbed Design

### *A. Design Rationale & Components Selection*

The objective of the experimental design is a reproducible, standards-aligned 5G O-RAN testbed that can compare three security configurations for the E2 interface: 1) no IPsec (i.e., a baseline), 2) classical IPsec using ECDH key exchange, and 3) IPsec using a NIST-standardised post-quantum KEM (ML-KEM) for IKEv2. The testbed should align with O-RAN alliance WG11 security specifications for interface protection and be built on widely available commercial-off-the-shelf (COTS) hardware and open-source components to enable reproducibility. Moreover, it must implement realistic RAN control and user-plane behaviour including Near-RT RIC xApp signalling and allows instrumentation of tunnel setup latency and steady-state resource metrics. Finally, it should allow straightforward substitution of KEM implementations (e.g., liboqs/strongSwan). These constraints will inform component selection for our designed testbed.

To balance realism and reproducibility, our design will utilise an open-source, containerisable software stack including srsRAN for RAN and UE emulation [10], Open5GS as the 5G Core [31], FlexRIC as the Near-RT RIC platform running representative xApps [12], and strongSwan (patched with liboqs) for IPsec/IKEv2 with both classical and PQC KEMs [13]. The choice of srsRAN/Open5GS/FlexRIC provides a standards-conformant E2 emulation while remaining deployable on commodity server hardware. Alternative options such as the Colosseum testbed [32], Free5GC [33], O-RAN Software Community (OSC) RIC [34] and the Liverpool 5G simulator [35]were explored but not selected based on pragmatic considerations, ensuring feasibility, reliability, and reproducibility within the scope of this work.

*B. Testbed Components*

An outline of the software components that collectively enable the deployment and evaluation are as follows.

**srsRAN –** In our testbed, O-RAN with its subcomponents and internal interfaces are based on srsRAN, and is a complete Open RAN solution developed by Software Radio Systems. srsRAN adheres to the 3GPP 5G system architecture by implementing functional splits between DU and CU. Furthermore, the CU is disaggregated into the Control Plane (CU-CP) and the User Plane (CU-UP), enabling greater flexibility and scalability in the network design, making it a complete O-RAN implementation [10].

**Open5GS –** Is an open-source implementation for the 5G Core, complying with 3GPP Release-17 specification. Documentation is available for building a 5G Core using Open5GS here [11]. Most notably, the proposed framework features a lightweight, containerised deployment of the Open5GS 5G Core, simplifying the development of the testbed and minimising the effort required for 5G Core deployment and integration.

**srsUE –** The srsRAN implementation does not include a UE implementation by default. Instead, the prototype 5G UE (srsUE) of srsRAN 4G can be used for the end-to-end test network [36]. srsUE is suitable for testing srsRAN in preliminary testing and proof-of-concept scenarios without using expensive hardware. In our testbed, neither UE devices nor hardware radio units are used. Instead of transmitting/receiving real RF signals using hardware modules, srsRAN implements an abstraction to emulate the radio channel between the UE and the gNB using ZeroMQ sockets.

**ZeroMQ –** Is a messaging library suitable for high-performance requirements of distributed or concurrent applications [37]. ZeroMQ's support for a range of common messaging patterns across multiple transport mechanisms enable it to be effectively used in radio channel emulation. This implementation makes it perfect for lab testing and development.

**FlexRIC –** Is an open-source near-RT RIC and an E2 node emulator agent. It has implementations of E2SM-KPM v2.01/v2.03/v3.00 and E2SM-RC v1.03 service models and includes several xApps compliant with O-RAN Alliance specifications. FlexRIC is developed with flexibility of development and deployment at its core [38]. It enables building specialised service-oriented controllers, while running its components and xApps in a set of orchestrated containers in isolation.

All these open-source components are widely used in the research community for evaluating the deployment and performance of O-RAN networks and therefore provide a reliable basis for our experiments.

*C. Topology Evolution Towards the Final Design*

The topology for our testbed had evolved through three

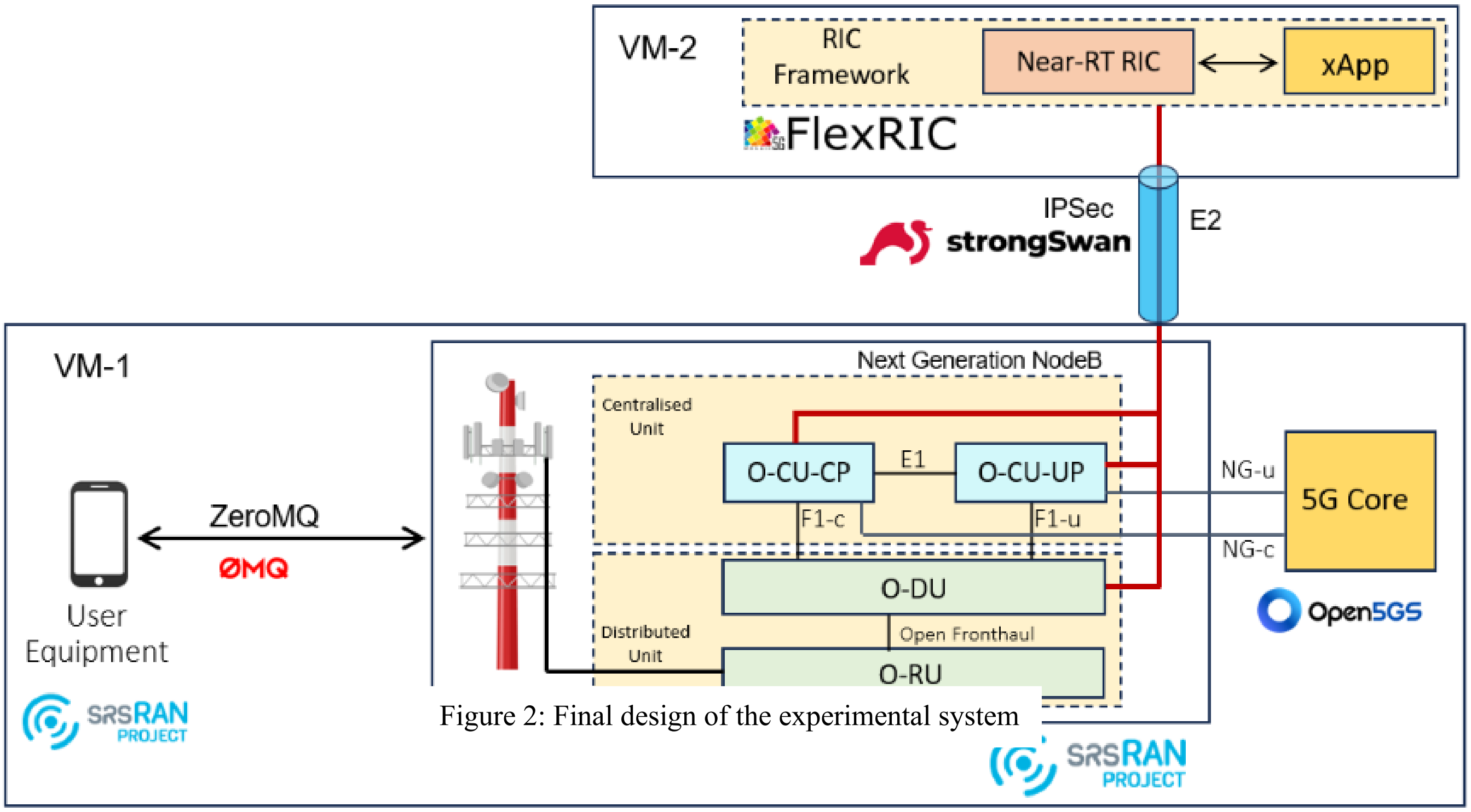

Figure 2: Final design of the experimental system

design stages:

1) Minimal functional topology (proof-of-concept). In this stage, a single host running srsRAN and Open5GS with a collocated RIC to validate basic E2 signalling and IKEv2/IPsec mechanics. This stage verified functional compatibility between RAN, RIC, and IPsec stacks and established measurement baselines.
2) Distributed software topology (isolation and instrumentation): functions were separated into distinct virtual machines/containers including (a) O-CU/O-DU (srsRAN), (b) 5G Core (Open5GS), (c) Near-RT RIC (FlexRIC + xApps), and (d) security gateway endpoints (strongSwan). Separating these roles enabled realistic network hops, latency characterisation, and per-node resource monitoring. It also allowed the IPsec tunnel endpoints to be placed in realistic locations. For instance, between the O-CU host and the Near-RT RIC host to emulate cross-domain E2 protection.
3) Final reproducible testbed as shown below in Fig. 2. The final design consolidates lessons from earlier stages into a modular topology suitable for repeatable experiments.

The key elements are:

a. O-RU / O-DU / O-CU (srsRAN split): srsRAN implements the gNB split (DU + CU functions). In the testbed, the logical gNB endpoint of E2 is hosted on the O-CU node.
b. Near-RT RIC (FlexRIC): Hosted on a separate node reachable over the emulated network. It runs representative xApps to generate realistic E2 control traffic and closed-loop signalling.
c. IPsec endpoints (strongSwan): Configured to protect the E2 interface between the O-CU and Near-RT RIC. The IKEv2 configuration can be toggled between classical (ECDH) and ML-KEM key exchanges. ML-KEM is provided by liboqs/strongSwan integration and uses NIST-recommended parameter sets for experimentation.
d. Core network (Open5GS) and UE traffic: Deployed to produce normal session activity and user plane load where required by specific experiments, but the E2 path is the primary measurement focus.
e. Measurement and orchestration hosts: Dedicated hosts and scripts for triggering tunnel setups, capturing packet traces, and measuring tunnel setup latency (from IKE_SA_INIT to IKE_AUTH completion and IPsec SA/ESP readiness). All nodes are time-synchronised (NTP) to ensure accurate latency measurement.

This final arrangement balances interface realism with experimental control, and it reflects WG11's recommendation to protect E2 using IPsec while allowing the research focus to be strictly on the cryptographic handshake overhead.

In Fig. 2, IPsec is realised using strongSwan with IKEv2. For classical experiments, an ECDH (e.g., curve25519) key exchange is used. For PQC experiments, the IKEv2 exchange is altered to use ML-KEM (CRYSTALS-Kyber / NIST ML-KEM) either as a pure KEM or in hybrid mode (classical + PQC). The implementation leverages liboqs support compiled into strongSwan's IKE daemon, allowing selection among ML-KEM parameter sets (512/768/1024) to study the performance trade-offs. The design keeps ESP protection (symmetric encryption) unchanged across configurations so that the measured differences focus on the key-establishment phase (IKE). The IPsec endpoint placement emulates a realistic cross-domain tunnel between the O-CU host and the Near-RT RIC host, consistent with WG11 guidance.

### *D. Workloads & Metrics*

Workloads are designed to stress both the control plane and the cryptographic handshake. First, E2 signalling workload: xApps on the Near-RT RIC generate periodic control messages and subscriptions to exercise E2 procedures at realistic rates to observe if tunnel establishment disrupts control loops. Secondly, UE/session traffic, uplink/downlink data flows are generated to validate that IPsec payload processing does not create steady-state overheads under PQC. In terms of measured metrics, we have identified the following. Tunnel setup latency measures the time between initiating IKEv2 and completion of IPsec Security Association (SA) establishment, which is measured from packet captures and IKE logs. xApp stability/control-loop impact that monitors message processing times and any error/recovery events at the Near-RT RIC.

Finally, the following instruments are used: tcpdump to capture packets at E2 endpoints, IKEv2 debug logs, system resource monitors, and automated scripts to repeat experiments across multiple runs for statistical confidence.

## IV. TESTBED IMPLEMENTATION

### *A. Hardware and Virtual Machines Configurations*

The implementation creates a multi-host software testbed using open-source components: srsRAN (gNB/UE), Open5GS (5G Core), FlexRIC (Near-RT RIC and xApps), and strongSwan (IPsec/IKEv2). The strongSwan instance is compiled with liboqs to support the NIST-standardised ML-KEM (CRYSTALS-Kyber) parameter sets. The testbed was deployed on a high-performance server with dual Intel Xeon Gold 6248 CPUs 2.50 GHz, 256 GB DDR4 ECC RAM, and VirtualBox v7.1.12 to support the creation of VMs. The following VMs were configured as shown in Fig. 3 below. VM1 has 2 vCPUs, 16GB memory, runs Ubuntu 22.04 LTS, and includes 2 network interfaces: 1) *enp0s8*, internal network only for interconnection; and 2) *enp0s9*, internal network only for management. VM2 has a similar configuration to VM1 except for 8GB memory. Finally, a third virtual machine, which is shown as "server" in Fig. 3, with the same configurations as that of VM2 was configured to act as the Certificate Authority (CA) for IPSec authentication.

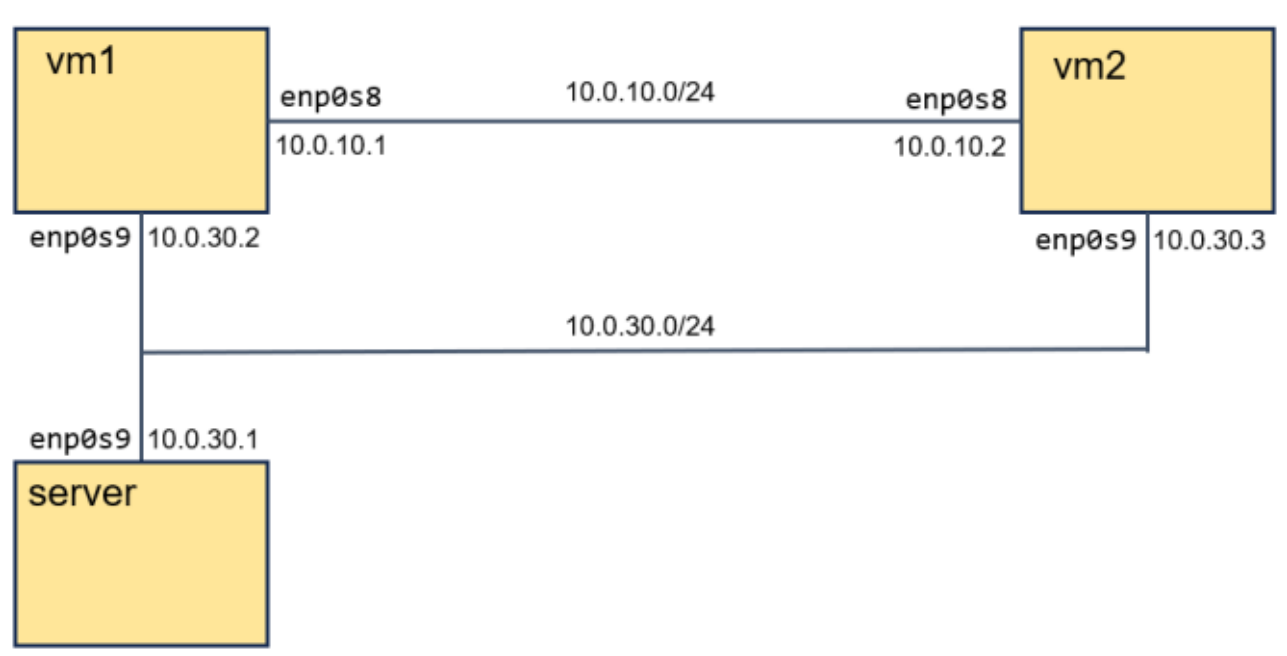

Figure. 3: VMs Configurations & Connectivity

### *B. Deployment of the 5G Network Components*

#### ***1. srsUE***

srsUE is part of srsRAN 4G, but the included experimental 5G UE can be used for our purposes. The installation can be done following these steps assuming dependencies are already installed.

```
# git clone https://github.com/zeromq/czmq.git
# make
# git clone https://github.com/srsRAN/srsRAN_4G.git
```

When compiling srsRAN 4G, ZMQ should be enabled.

```
/srsRAN_4G/build#  cmake  ../  -DENABLE_EXPORT=ON  -
DENABLE_ZEROMQ=ON
/srsRAN_4G/build# make
/srsRAN_4G/build# make test
```

The last command verifies the installation by running the built-in test. Upon successful installation, the executable file `srsue` should be available in `srsRAN_4G/build/srsue/src/`.

#### ***2. srsRAN***

After installing the build tools and mandatory requirements, we clone the git repository

```
# git clone
https://github.com/srsRAN/srsRAN_Project.git
```

ZeroMQ is disabled by default, so it should also be enabled here when building as follows:

```
# cmake ../ -DENABLE_EXPORT=ON -DENABLE_ZEROMQ=ON
# make -j $(nproc)
# make install
```

#### ***3. Open5GS***

The srsRAN Project provides a containerised deployment of the Open5GS core network, a docker container, significantly simplifying the process of setting up a functional 5G core. Assuming Docker compose is installed, the 5G core included in srsRAN project can be built as follows:

```
/srsRAN_Project/docker# docker compose --build 5gc
/srsRAN_Project/docker# docker compose start 5gc
[+] Running 1/1
    Container open5gs_5gc Started
```

#### ***4. FlexRIC***

This stage involves the installation and configuration of FlexRIC followed by verification of xApp integration to ensure correct operation and interaction with the deployed 5G network components. These steps establish the foundation for subsequent performance evaluations involving secure E2 interface communication. After installing the necessary dependencies, we install FlexRIC as follows.

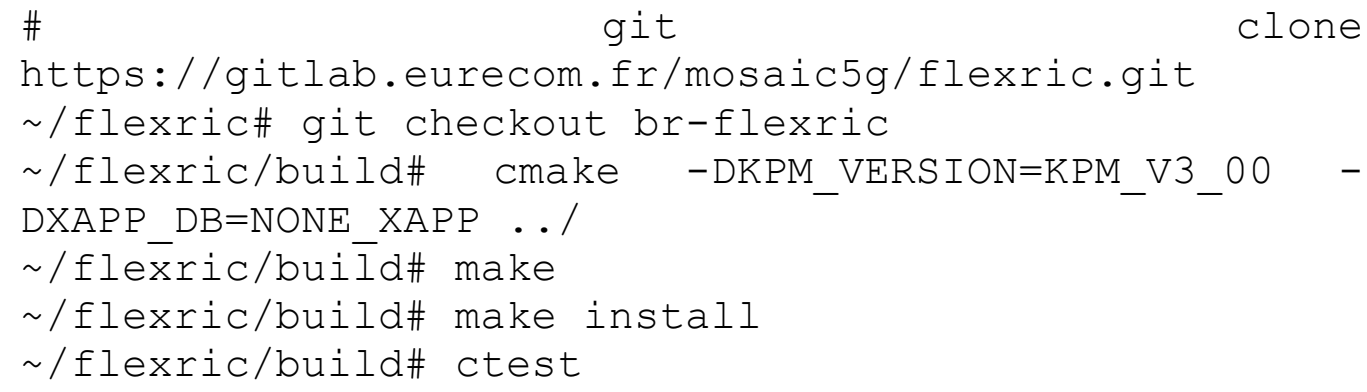

```
#                    git                    clone
https://gitlab.eurecom.fr/mosaic5g/flexric.git
~/flexric# git checkout br-flexric
~/flexric/build#   cmake   -DKPM_VERSION=KPM_V3_00   -
DXAPP_DB=NONE_XAPP ../
~/flexric/build# make
~/flexric/build# make install
~/flexric/build# ctest
```

The last command will run the built-in test to verify the installation. Upon successful implementation, the executable file, `nearRT-RIC`, should be available in `flexric/build/examples/ric`.

The final setup is shown in Fig. 4 below with established routes between gNodeB and RIC.

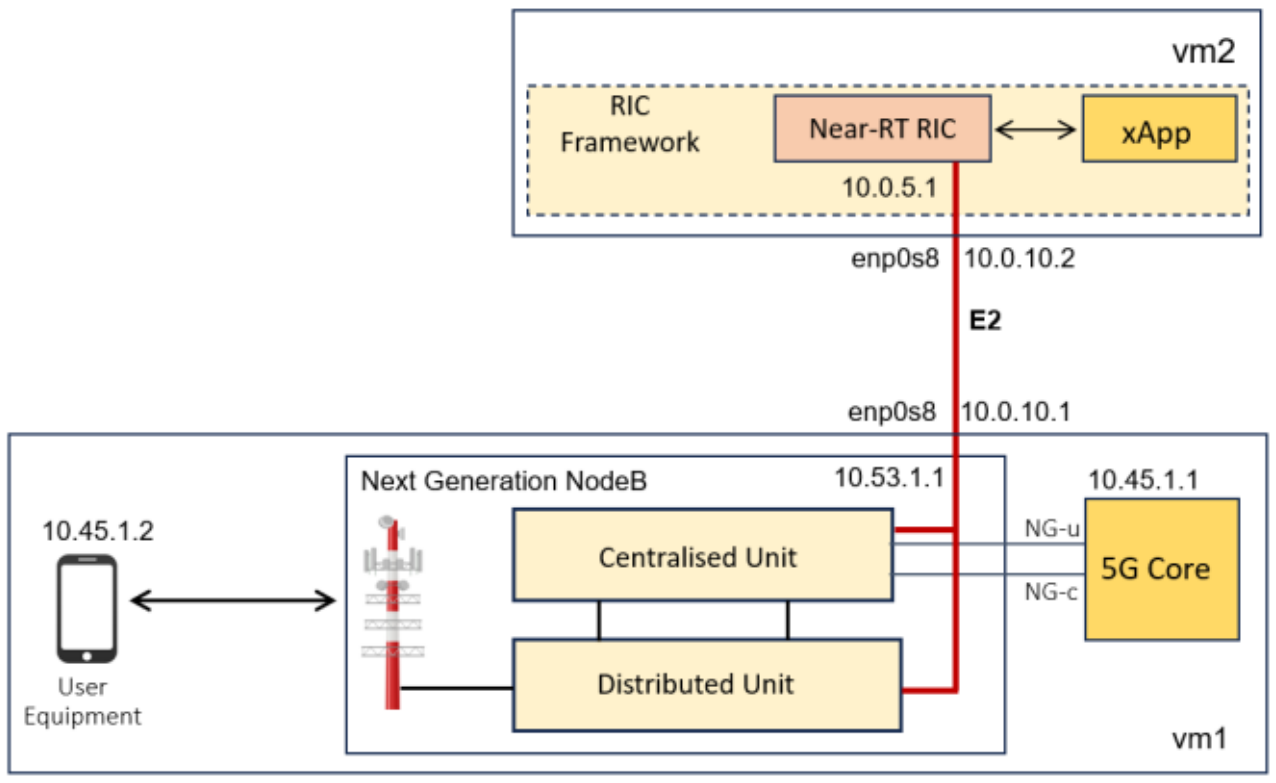

Figure. 4: The final setup of the experimental system

### *C. Implementation of Interface Security*

In this section, we explain the integration of PQC into the O-RAN E2 interface using strongSwan with ML-KEM support. Note that support for PQC algorithms in strongSwan is available only from version 6.0.0 onwards. After installing dependencies, the latest source code of strongSwan 6.0.1 is installed via its website: https://download.strongswan.org/strongswan-6.0.1.tar.bz2. While preparing for building, it is important to ensure that `--enable-ml` is used as an option to prepare the installation with ML-KEM cryptographic primitives. After compiling and installing strongSwan, the installation and ML-KEM algorithm support can be verified as shown below in Fig. 5.

```
root@vm1:~# swanctl --version
strongSwan swanctl 6.0.1
root@vm1:~# which charon-systemd
/usr/sbin/charon-systemd
root@vm1:~# swanctl --list-algs | grep -Ei kem
  ML_KEM_512[ml]
  ML_KEM_768[ml]
  ML_KEM_1024[ml]
root@vm1:~#
```

Figure. 5: strongSwan version with ML-KEM support

Based on the two VM setup, the interconnecting link IP address plans and the subnets used in O-RAN and RIC can be respectively used for designing the IKE/IPSec policies for SAs. Configuration parameters for IPSec were as follows. ESP is used as the method to encrypt payloads for data secrecy and mode is set to tunnel mode. Authentication is performed using PKI, where the VM "server", as shown in Fig. 3, is used as the CA. Encryption uses AES-256-GCM, KEM is ECDH over Curve25519, and integrity is verified using prfsha256. Once all the certificates are generated and signed, the configuration file `/etc/swanctl/swanctl.conf` is updated as per the designed IPSec policy. Below is an example of the IPSec configuration file of VM1.

```
connections {
   e2 {
      proposals = aes256gcm16-prfsha256-curve25519
      remote_addrs = 10.0.10.2
      local {
       auth = pubkey
       certs = vm1Cert.pem
     }
     remote {
       auth = pubkey
       id = "C=CH, O=strongswan,
CN=vm2.strongswan.org"
     }
     children {
       app1 {
         local_ts = 10.0.10.1/24, 172.16.1.0/27
         remote_ts = 10.0.10.2/24, 172.16.2.0/27
         esp_proposals = aes256gcm16-prfsha256-
curve25519
         start_action = trap
       }
     }
   }
}
```

Note that we had to change the IP address of the Near-RT RIC to run on 172.16.2.34 because, by default, both gNB and Near-RT RIC are configured to run on a single VM (i.e., the RIC runs on 127.0.0.1 by default).

### D. Execution & Validation of the Testbed

The network components should be started in the following sequence to ensure proper operation.

**1. 5G Core**

Initialise 5G core using docker:

```
docker compose up 5gc
```

**2. Near-RT RIC**

Run the FlexRIC executable:

```
~/flexric/build/examples/ric#./nearRT-RIC
```

**3. O-RAN (gNB)**

Run the `gnb` executable, with the configuration file specific for E2 settings.

```
../srsRAN_config# gnb -c gnb_zmq_e2.yaml
```

Upon starting the gNB, the following can be observed as shown in Fig. 6:

- The "Connection to AMF on 10.53.1.2:38412" message indicates that the gNB initiated a connection to the 5G core.
- The "Connection to NearRT-RIC on 172.16.2.34:36421" message indicates that the gNB initiated a connection to the NearRT-RIC sucessfully.

```
root@vm1:~/project-lab/srsRAN_config# ls
gnb_zmq_e2.yaml  gnb_zmq.yaml  ue_zmq.conf  ue_zmq_e2.conf
root@vm1:~/project-lab/srsRAN_config# gnb -c gnb_zmq_e2.yaml

--== srsRAN gNB (commit d90cd4e26d) ==--

Lower PHY in executor blocking mode.
Available radio types: zmq.
Cell pci=1, bw=10 MHz, 1T1R, dl_arfcn=368500 (n3), dl_freq=1842.5 MHz, dl_ssb_arfcn=368410, ul_freq=1747.5 MHz

N2: Connection to AMF on 10.53.1.2:38412 completed
E2AP: Connection to Near-RT-RIC on 172.16.2.34:36421 completed
==== gNB started ===
Type <h> to view help
```

Figure 6: Successful initiation of the gNB

In the 5G core terminal, the gNB attachment is observed as shown in Fig. 7 below.

```
1] (../lib/sbi/nnrf-handler.c:938)
open5gs_5gc  | 08/20 17:43:56.265: [sbi] INFO: (NRF-notify) NF registered [7b40df40-7ddc-41f0-9fb9-4fe0a88fb19c:1] (../lib/sbi/nnrf-handler.c:924)
open5gs_5gc  | 08/20 17:43:56.265: [sbi] INFO: [UDR] (NRF-notify) NF Profile updated [7b40df40-7ddc-41f0-9fb9-4fe0a88fb19c:1] (../lib/sbi/nnrf-handler.c:938)
open5gs_5gc  | 08/20 17:43:56.267: [sbi] INFO: [7b40df40-7ddc-41f0-9fb9-4fe0a88fb19c] NF registered [Heartbeat:10s] (../lib/sbi/nf-sm.c:221)
open5gs_5gc  | 08/20 17:43:56.268: [nrf] INFO: [7b420708-7ddc-41f0-8f03-45d597893cea] Subscription created until 2025-08-21T17:43:56.268736+02:00 [validity_duration:86400] (../src/nrf/nnrf-handler.c:445)
open5gs_5gc  | 08/20 17:43:56.269: [sbi] INFO: [7b420708-7ddc-41f0-8f03-45d597893cea] Subscription created until 2025-08-21
open5gs_5gc  | 08/20 18:06:10.833: [amf] INFO: gNB-N2 accepted[10.53.1.1]:55505 in ng-path module (../src/amf/ngap-sctp.c:113)
open5gs_5gc  | 08/20 18:06:10.833: [amf] INFO: gNB-N2 accepted[10.53.1.1] in master_sm module (../src/amf/amf-sm.c:741)
open5gs_5gc  | 08/20 18:06:10.843: [amf] INFO: [Added] Number of gNBs is now 1 (../src/amf/context.c:1231)
open5gs_5gc  | 08/20 18:06:10.843: [amf] INFO: gNB-N2[10.53.1.1] max_num_of_ostreams : 30 (../src/amf/amf-sm.c:780)
```

Figure 7: Successful attachment of gNB to 5G core

**4. User Equipment**

First, we need to create a namespace for the UE

```
~/project-lab# ip netns add ue1
```

Then, we initiate the UE by running the executable, with the configuration file modified with configuration settings for ZMQ-based RF driver and E2.

```
#./project-lab/srsRAN_4G/build/srsue/srcsrsue /root/project-lab/srsRAN_config/ue_zmq_e2.conf
```

Once the srsUE is successfully attached, a message should appear on the 5G Core console. Finally, after configuring routing among these components, traffic between the UE and 5G Core is generated using ipref3 to validate the experimental setup. Note that we ran the iperf3 server on the 5G core as a docker container, while the iperf3 client is running in the UE network namespace.

**5. xApp**

To verify the proper working of the experimental system, we ran an xApp available with the FlexRIC installation. The xApp `xapp_oran_moni` connects to the Near RT-RIC and uses E2SM_KPM service modules to subscribe for measurement data. The metric names to be passed to the xApp are in the srsRAN configuration file, which was updated with the Near RT-RIC IP address as shown below in Fig. 8.

```
root@vm2:~/project-lab/flexric/build/examples/xApp/c/monitor# cat xapp_mon_e2sm_kpm.conf
SM_DIR = "/usr/local/lib/flexric/"

# supported name = xApp
Name = "xApp"
NearRT_RIC_IP = "172.16.2.34"
E42_Port = 36422
```

Figure 8: xApp config file modified with near RT-RIC

## V. Evaluation and Results Analysis

In this section, we present measurement-based evaluation of the performance impact of integrating PQC KEM into IPsec protecting the O-RAN E2 interface.

### *A. Performance Metrics*

Before we dive into the experiments' configuration, we provide details of the three primary metrics we chose to focus on in this evaluation. For each metric, we develop a test case and configuration to measure it effectively.

1. **Latency introduced by IPsec**. This provides a direct measure of the overhead imposed on traffic between two endpoints due to securing an O-RAN interface. It is particularly relevant for latency-sensitive applications using the E2 interface.
2. **IPsec Security Association (SA) setup time**. This measures the impact of using PQC KEM algorithms in comparison to classical KEM, as an additional overhead. It reflects how PQC adoption may impact session initiation delays in practice.
3. **Latency with IPSec SA setup for xApp**. This measures the latency on xApp traffic during the IPSec SA setup. It reflects how PQC adoption may impact xApp's operation during the shorter period of session initiation.

To conduct a controlled experiment, and isolate the performance impact of different cryptographic primitives, the testbed was systematically configured for the following scenarios 1) baseline operation without IPSec, 2) IPsec with classical KEM, and 3) IPsec with PQC KEM.

### *B. Experimental Scenarios Setup*

#### ***1. Latency introduced by IPSec***

The test setup consists of two VMs running Ubuntu (version 24.04.2 LTS) as shown below in Fig. 9. sockperf is used as the latency testing tool. sockperf [39] is a utility designed for testing performance (latency and throughput) for network benchmarking.

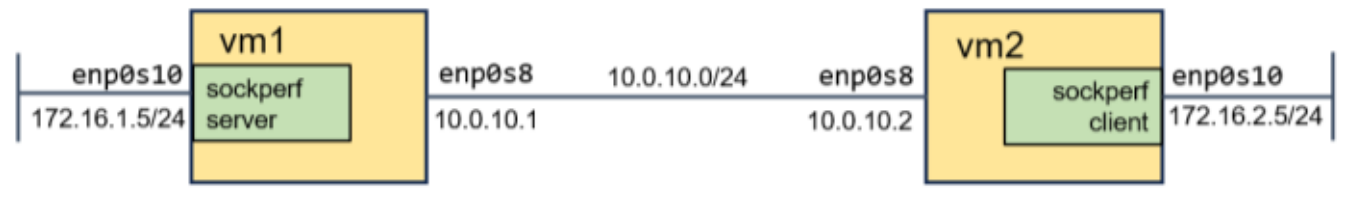

Figure. 9: VM connectivity for network latency observation

As shown in Fig. 9, the sockperf server is running on VM1, while the client is running on VM2. Here, we run 100 iterations of the following command with and without IPSec:

```
sockperf ul --ip "$SERVER_IP" --client_ip "$CLIENT_IP" > "$OUTPUT_DIR/iteration-$i.txt"
```

#### ***2. IPSec SA Setup Time***

In this test case, the time taken to setup an IPsec tunnel established between two VMs is measured to evaluate whether the use of different KEM leads to variations in tunnel setup time. We used the same setup as in Fig. 9. However, we considered tcpdump to analyse the timestamps of IKEv2 messages exchanged during the initiation process. After that, we ran the following test cases: 1) IPSec setup with classical KEM (ECDH was used in the IKE SA proposal namely aes256gcm16-prfsha256-curve25519), 2) IPSec setup latency with ML-KEM-768 (ML-KEM was used with aes256gcm16-prfsha256-mlkem768), and 3) IPSec setup latency with ML-KEM-1024 (ML-KEM was used with aes256gcm16-prfsha256-mlkem1024).

#### ***3. Latency with IPSec SA setup for xApp***

In this test case, the time between the execution of an xApp and the first packet leaving the VM interface was measured to evaluate whether the use of ML-KEM leads to a delay in the start of control traffic flow. Using the same setup in Fig. 9, we used tcpdump to analyse the timestamps of xApp control messages (SCTP) exchanged. On VM2, we started packet capturing `tcpdump -i enp0s8 -c 5`. Then, on VM1, the xApp was initiated as follows:

```
/root/project-lab/flexric/build/examples/xApp/c/monitor/xapp_oran_moni -c xapp_mon_e2sm_kpm.conf
```

Two test cases were evaluated here. First, IPSec was configured between VM1 and VM2, with ECDH as the KEM. For the second case, IPSec was configured with ML-KEM1024 as the KEM.

### *C. Results and Analysis*

After running the test cases illustrated in the previous sub-section, we collected performance data from the results files to analyse them in this paper. We first analyse the latency introduced by IPSec. As can be seen in Fig. 10 and Table 1, it is evident that in comparison to the case of baseline latency, IPSec with AES256 introduces an additional latency of around 165 μs attributable to encryption. This is an expected result, as encryption of data involves additional processing for transforming plaintext into ciphertext. This processing causes an additional latency on top of the network latency applicable in the baseline case.

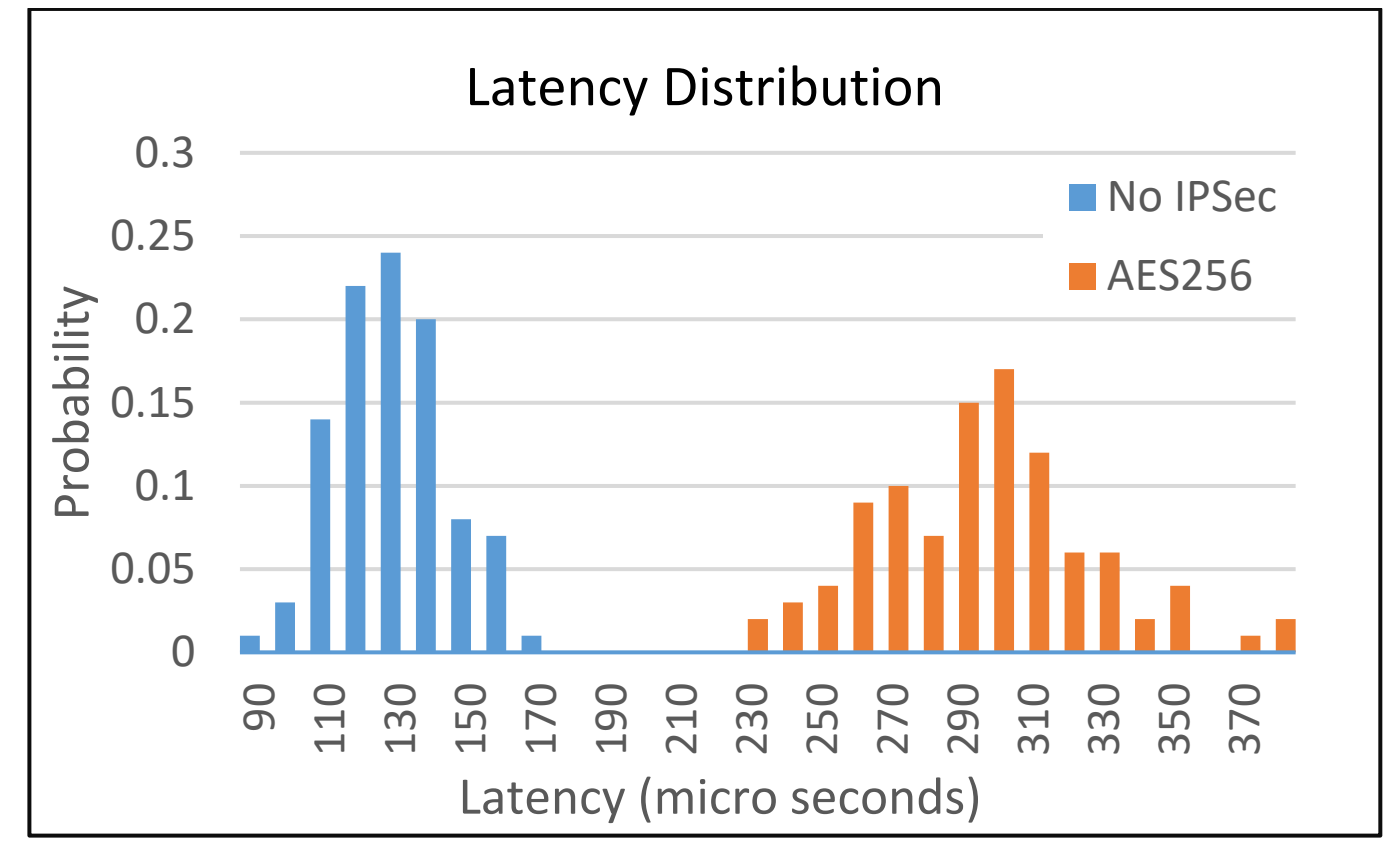

Figure 10: Latency distribution – No IPSec vs. IPSec (AES 256)

Table 1: Latency Observations with/without IPSec

| Measurement | No IPSec | IPSec with AES256 |
| --- | --- | --- |
| Average latency (μs) | 125 | 290 |
| Median latency (μs) | 125 | 291 |

In terms of IPSec SA setup time, we calculated *ike_init*, *ike_auth* and *child_sa* setup times extracted from packet header timestamps in each result file.

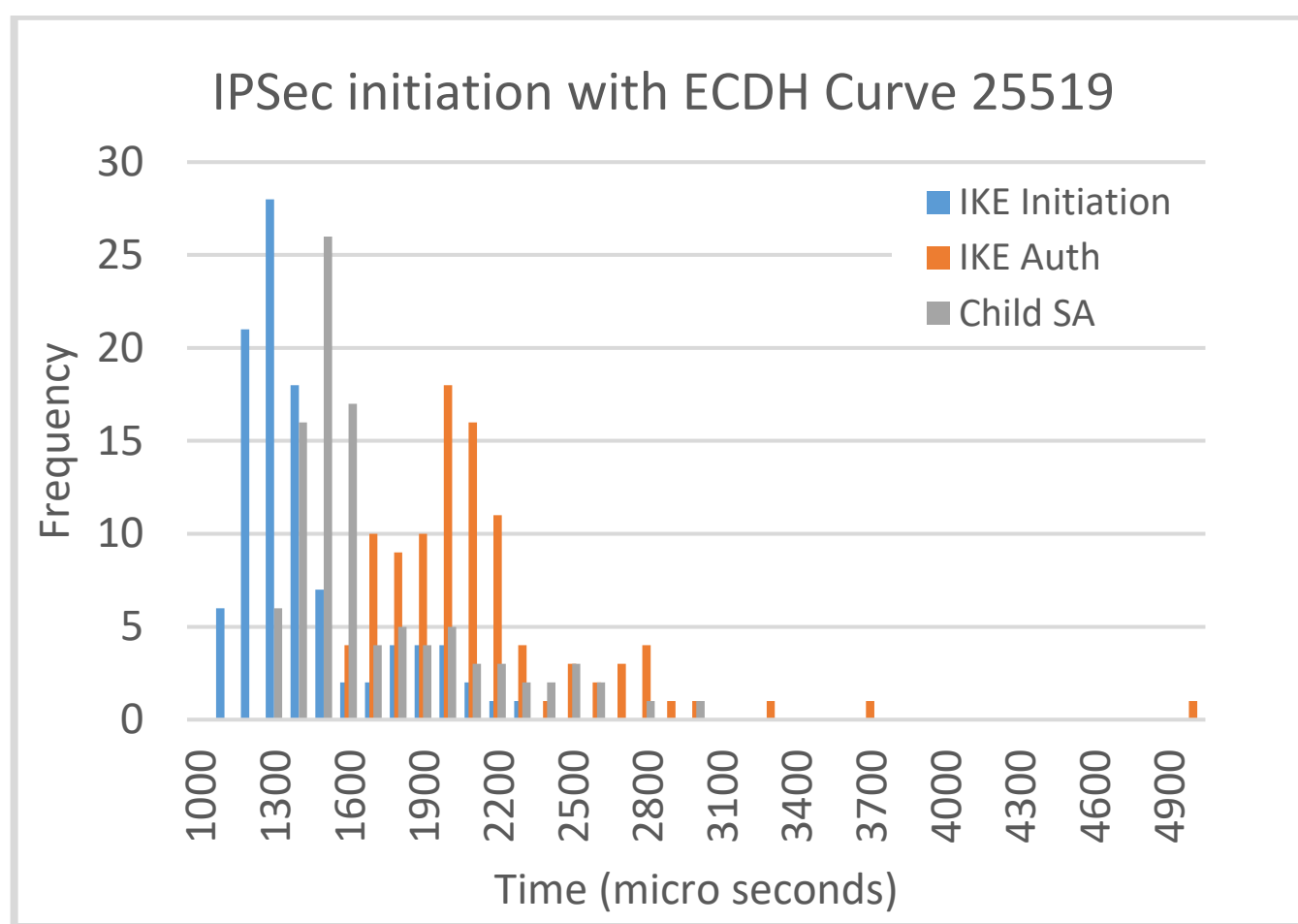

Figure 11: Initiation times with ECDH Curve 25519

Table 2: Initiation times with ECDH Curve 25519

| Measurement | IKE Initiation | IKE Authorisation | Child SA setup |
|---|---|---|---|
| Average time (µs) | 1363 | 2093 | 1658 |
| Median time (µs) | 1284 | 1998 | 1508 |

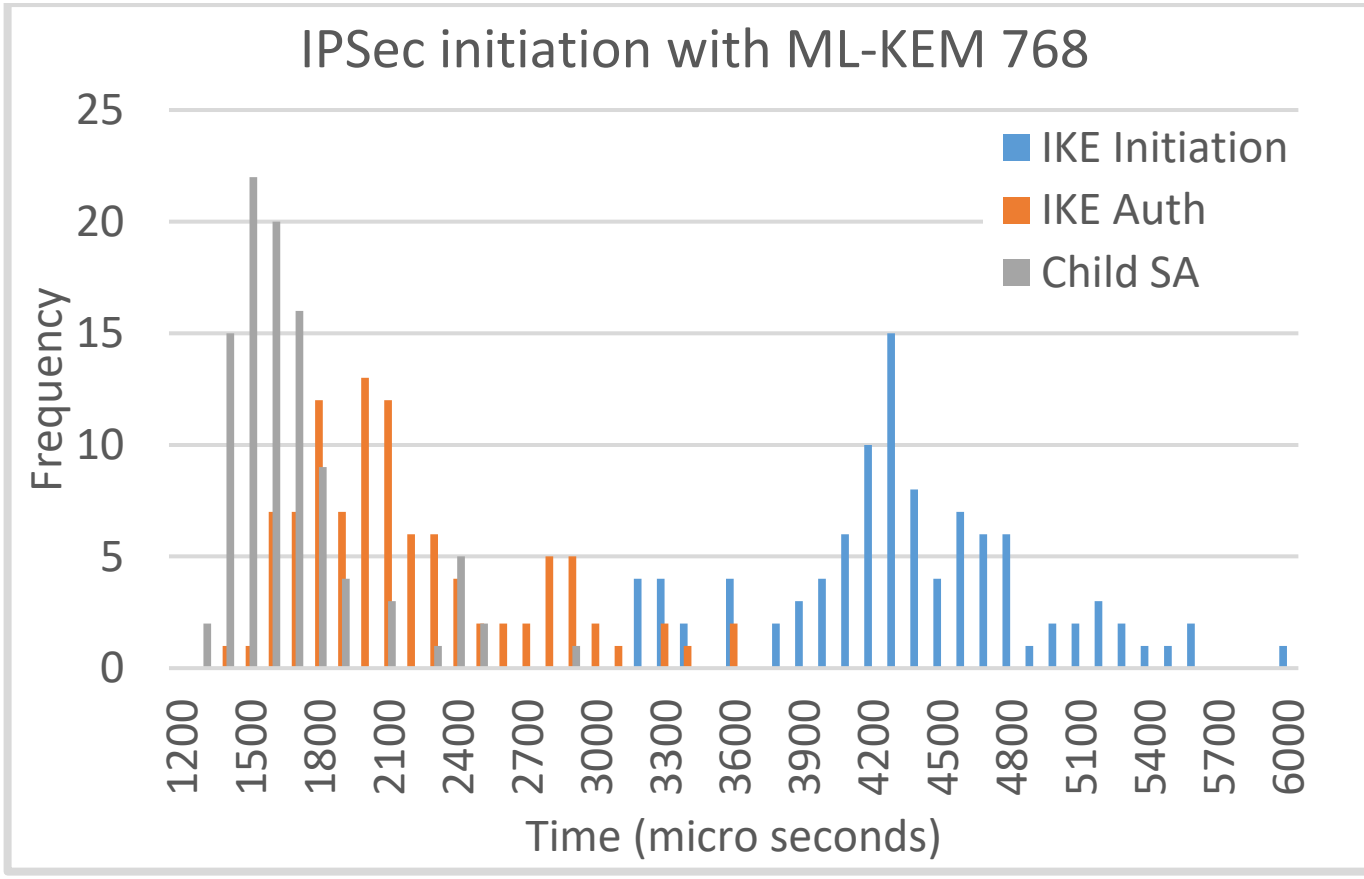

Figure 12: Initiation times with ML-KEM 768

Table 3: Initiation times with ML-KEM 768

| Measurement | IKE Initiation | IKE Authorisation | Child SA setup |
|---|---|---|---|
| Average time (µs) | 4306 | 2141 | 1633 |
| Median time (µs) | 4281 | 2020 | 1548 |

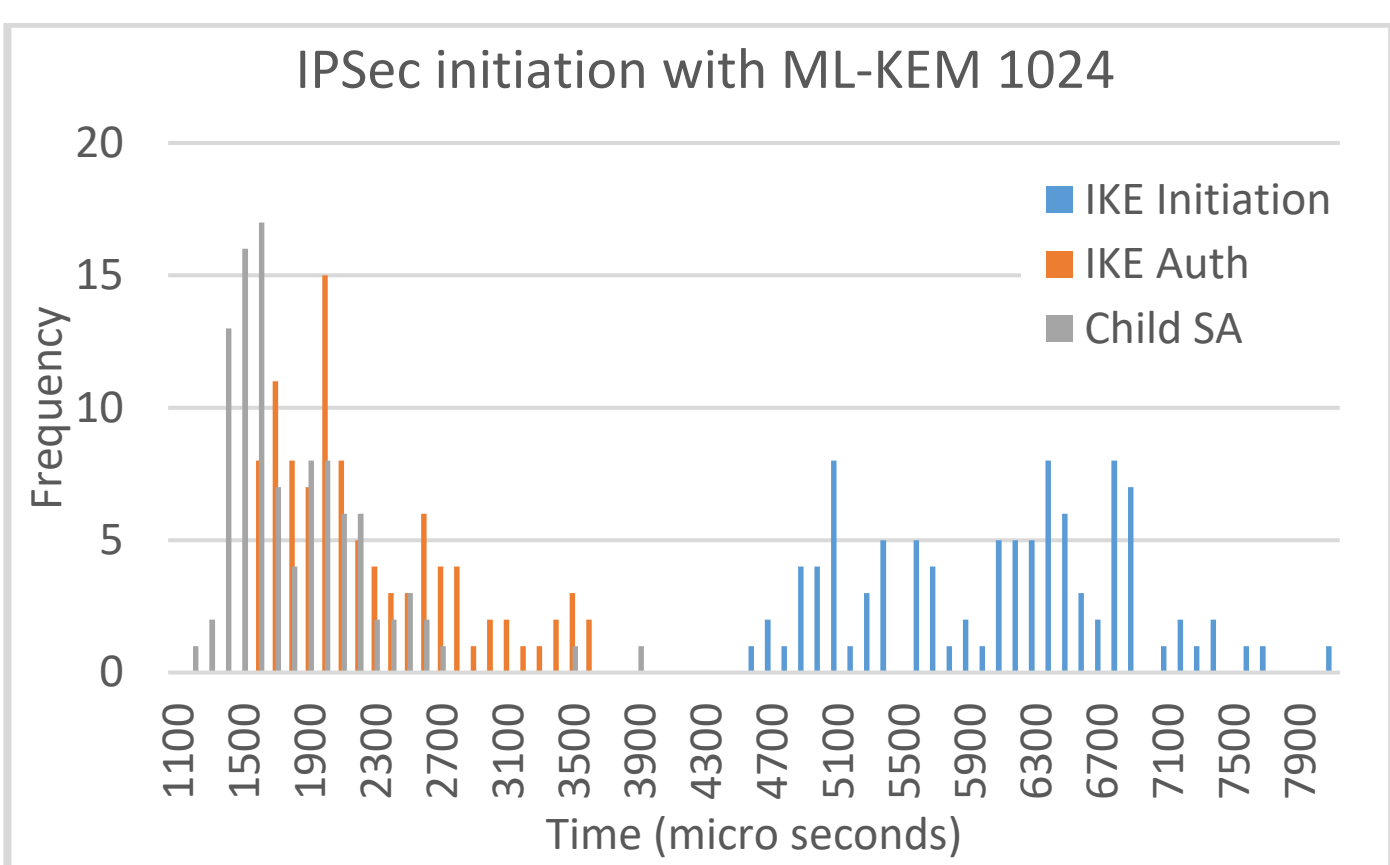

Figure 13: Initiation times with ML-KEM 1024

Table 4: Initiation times with ML-KEM 1024

| Measurement | IKE Initiation | IKE Authorisation | Child SA setup |
|---|---|---|---|
| Average time (µs) | 6026 | 2198 | 1768 |
| Median time (µs) | 6146 | 2014 | 1649 |

In comparison to the classic KEM ECDH Curve 25519 as the baseline, IPSec with ML-KEM-768 and ML-KEM-1024 introduces an additional delay of 3 ms and 4.7 ms respectively for the IKE initiation process. For ML-KEM-768 and ML-KEM-1024 this is an expected result, as for larger key sizes more computations are required. Most importantly, this provides evidence for the increased SA setup time for ML-KEM in comparison to the classical KEM (ECDH Curve 25519) baseline.

Finally, we calculate the time between the execution of the xApp command and the first xApp packet leaving VM2 interface (i.e., RIC) using the xApp packet flow for each iteration, which was captured by tcpdump. In both cases, with and without IPSec, we got the same results as shown in Fig. 14 below. This means IPSec setup has no impact on xApp performance. This is expected, since IPSec should be operational before any data is transferred between gNB and the near real-time RIC.

```
mario@vm2:~/project-lab/test-cases_6.2.4$ ./delay.sh
xApp start time   : 1756279277.158345
First packet time: 1756279279.191873
Delay (seconds)   : 2.033528
```

Figure 14: Delay between xApp command and first packet leaving the RIC interface

## VI. Conclusion & Future Work

In this paper, we presented a testbed-based evaluation of integrating a NIST-aligned module-lattice KEM (ML-KEM / CRYSTALS-Kyber) into IKEv2/IPsec protecting the 5G O-RAN E2 interface. Using an open-source stack (srsRAN, Open5GS, FlexRIC) and a liboqs-enabled strongSwan, we compared three configurations: no IPsec, classical ECDH-based IPsec, and ML-KEM-based IPsec, under realistic Near-RT RIC signalling workloads. Repeated, automated trials show that ML-KEM primarily affects the key-establishment phase where tunnel setup latency increases by a small, reasonable ≈3~5 ms relative to classical IPsec, while the runtime behaviour of Near-RT RIC xApps and control-loop operation remained stable in our experiments. These results indicate that ML-KEM–based IPsec on the E2 interface is practically feasible on contemporary server-class hardware and can form part of early, staged quantum-safe migration strategies for Open RAN deployments.

For future work, we are pursuing two main directions. First, extending the experimental framework to protect O-RAN interfaces that rely on TLS (e.g., A1, O1, O2) using PQC and hybrid PQC–classical key exchange. This will evaluate the impact of ML-KEM and alternative NIST-standardised KEMs on TLS 1.3 handshake latency, session resumption, and control-plane message delay under realistic RAN management and orchestration workloads. Secondly, despite the many advantages of PQC, it cannot, on its own, address all emerging security challenges. Consequently, a key future research direction of this work is the investigation of how Federated Learning (FL) can be leveraged to enhance PQC security. Since O-RAN environments comprise modular components often operated by different vendors, it makes it ideally suited to FL, enabling collaborative learning for PQC security optimisation across distributed entities without the need to share sensitive cryptographic data or rely upon a centralised learning framework. We aim to explore how FL could be used to optimise PQC security to maintain quantum-resistant implementations. Initial focus will be on:

- Optimising PQC: Leveraging federated feedback to establish effective algorithm selection (e.g. lattice-based, hash-based) and parameter settings (e.g. key size, error distributions), which achieve a balance between security and efficiency.
- Collaborative anomaly detection: Locally trained models provide insights that are aggregated into federated updates, capturing behavioural, session establishment or implementation anomalies (e.g. leakage, timing), without exposing raw data.
- Post deployment hardening: As PQC systems face evolving threats, FL enables collaborative and rapid identification of novel attacks (e.g. side-channel, downgrade) and implementation weaknesses based on federated insights from live systems, supporting ongoing attack resilience.